\documentstyle[12pt]{article}

\title{The Leading Order of the Theory of Strong Perturbations in Quantum
      Mechanics}

\author{Marco Frasca\ddag \\
Via E. Gattamelata,3 \\
00176 Roma (Italia)}

\date{}
\begin{document}
\maketitle

\newpage

\begin{abstract}
We prove that, for a quantum system that undergoes a strong perturbation,
the solution of the leading order equation of the strong field
approximation
(M.Frasca, Phys. Rev. A, {\bf 45}, 43 (1992)) can be derived by the
adiabatic approximation. In fact, it is shown that greatest is the
perturbation and more similar the quantum system is to an adiabatic
one, the solution being written as a superposition of eigenstates of the
time-dependent perturbation.A direct consequence of this result is that
the solution of the Schr\"{o}dinger equation in the interaction picture,
in the same approximation for the perturbation, coincides with the one of
the
leading order of the strong field approximation. The limitation due to the
requirement that the perturbation has to commute at different times is so
overcome. Computational difficulties could arise to go to higher orders.
Beside, the method is not useful for perturbations that are constant in
time.
In such a case a small time series is obtained, indicating that this
approximation is just an application to quantum mechanics of the
Kirkwood-Wigner expansion of statistical mechanics. The theory obtained
in this way is applied to a time-dependent two-level spin model, already
considered for the study of the Berry's phase, showing that a geometrical
phase could arise if a part of the hamiltonian is considered as a strong
perturbation. No adiabatic approximation is taken on the parameters of the
hamiltonian, while their cyclicity is retained.
\end{abstract}

\newpage

\section{Introduction}

In a series of papers, I proposed a new perturbation approach to cope with
quantum systems that experience the effect of a large time-dependent
perturbation [1]. The main computational limitation that appears is
originating from the leading order equation that can be written as
${\displaystyle V(t)|\psi> = i\hbar\frac{d|\psi>}{dt}}$, being $V(t)$ the
perturbation. A general solution for this equation does not exist unless
we
take ${\displaystyle [V(t), V(t')] = 0}$. This condition appears to limit
the
usefulness of the method to some simple systems. Actually, the situation
is
a little more favourable than it does not seem at a first view.
We will see in a moment that the main trick to derive the perturbation
series
in [1], i.e. rescaling the time variable, can change the situation.

In fact, the leading order equation should be more correctly written as
\begin{equation}
    \lambda V(t) |\psi(t)> = i\hbar \frac{d|\psi(t)>}{dt} \label{eq:lo}
\end{equation}
being $\lambda$ the parameter taken to run away to infinity as in [1]. In
such
conditions we are able to find, in any case except computational
difficulties,
an approximate solution to the above equation,
that agrees with the strong field approximation proposed in
[1], as the above equation is identical to the one of ref.[2] to prove
the adiabatic theorem of quantum mechanics.
To see that this is indeed the case, with the rescaling proposed in
[1] of the time variable, i.e. $t\rightarrow\lambda t = \tau$, one gets
\begin{equation}
    V(\frac{\tau}{\lambda}) |\psi(\frac{\tau}{\lambda})> =
    i\hbar \frac{d|\psi(\frac{\tau}{\lambda})>}{d\tau}
\end{equation}
that shows as $\lambda$, the perturbation sthrength,
is a natural time-scale for the change of both the hamiltonian
and the wave-function. For very large $\lambda$, the variation is very
slowly and we can invoke the adiabatic approximation.

In the following we will use the no-rescaled equation as, in such a way,
we are able to see directly the approximation involved. The main physical
point
is that the stronger the perturbation, the more indistinguishable the
quantum
system is from an adiabatic one. This result also gives a general
approach to cope with equations in the form
${\displaystyle \lambda L(t) u(t) = \frac{du(t)}{dt}}$, being $\lambda$ a
very large parameter and $L(t)$ an
operator depending on the parameter $t$ and acting on the
vector $u(t)$. To our knowledge, in literature, the adiabatic theorem
has never been presented from this point of view.

Some computational difficulties could arise when one tries to go to higher
orders. This problem originates from the fact that the eigenstates of the
perturbation are time-dependent. We hope to treat this limitation in a
future paper, but we are however beyond the strong limitation discussed
above.

An important case is the one of a time-independent perturbation. Here we
face a difficulty of the method as we are able to get just a small time
series. This difficulty is simply indicating that we are applying the
high-temperature Wigner-Kirkwood method of statistical mechanics to
quantum systems. However, such kind of limitations are typical of
perturbation methods as one can see from the standard small-perturbation
approach applied to a simple two-level model.

The paper is so structured. In sec.2 we present the derivation of the
adiabatic theorem for strongly perturbed quantum systems and show that, in
the interaction picture, the solution, for a large perturbation, can be
written in the same form as for the strong field approximation. In sec.3
we consider the case of a time-independent perturbation showing that here
we face a small-time development. In sec.3 we apply our result to a
two-level model showing that if a part of the hamiltonian can be
considered
as a large perturbation then a Berry's phase arises
as computed in [3] for the adiabatic analog of the perturbation part of
the
considered system.

\section{A Derivation of the Adiabatic Approximation from a New Point of
View}

For our aims we consider the following Schr\"{o}dinger equation
\begin{equation}
    \lambda H(t)|\psi> = i\frac{d|\psi>}{dt} \label{eq:schroed}
\end{equation}
having $\lambda\rightarrow\infty$ and $H(t)$ a time-dependent hamiltonian
typical of the considered quantum system. Here and in the following we
set $\hbar = 1$. In order to show clearly the
approximations involved, we do not operate the rescaling
$t\rightarrow\lambda t$.
Instead, let us make the {\sl ansatz} as in the adiabatic approximation
\begin{equation}
    |\psi> = \sum_n c_n(t) e^{i\gamma_n(t)}
                           e^{-i\lambda\int_{t_0}^{t} E_n(t') dt'} |n; t>
                           \label{eq:ansatz}
\end{equation}
being
\begin{equation}
    H(t) |n; t> = E_n(t) |n; t>
\end{equation}
and
\begin{equation}
    \dot{\gamma}_n = <n; t|i\frac{d}{dt}|n; t>.
\end{equation}
The probability amplitudes, $c_n(t)$ are to be found. By a direct
substitution of eq.(\ref{eq:ansatz}) into eq.(\ref{eq:schroed}) one gets
\begin{equation}
    \dot{c}_m(t) = -\sum_{n \neq m} e^{i[\gamma_n(t) - \gamma_m(t)]}
                    e^{-i\lambda\int_{t_0}^t \Omega_{nm}(t') dt'}
                    <m; t|\frac{d}{dt}|n; t> c_n(t) \label{eq:cn}
\end{equation}
that is the sought equation to find the probability amplitudes. We have
set
$\Omega_{nm}(t) = E_n(t) - E_m(t)$. The standard approach to obtain a
set of solutions to eq.(\ref{eq:cn}) is to put it in integral form and
iterate by taking, at the leading
order, the probability amplitudes at the initial time $t_0$. In the limit
$\lambda\rightarrow\infty$ we have again the adiabatic approximation, that
is
$c_m(t)\approx c_m(t_0)$.

In fact, at the next order, one has to evaluate the integral
\begin{equation}
    I_{nm}(t) = \int_{t_0}^t
                 e^{i[\gamma_n(t') - \gamma_m(t')]}
                 e^{-i\lambda\int_{t_0}^{t'} \Omega_{nm}(t'') dt''}
                 <m; t'|\frac{d}{dt'}|n; t'> dt'
\end{equation}
that, in the limit $\lambda\rightarrow\infty$, has a strongly oscillating
exponential. In such a case we recognize the same situation as in ref.[2]
for the adiabatic theorem
and we have that the integral goes to 0 at least as
$\frac{1}{\sqrt{\lambda}}$, if the energy levels cross (for a more
rigourous mathematical approach to the asymptotic evaluation of integrals
we refer back to refs.[4]). Then, we can conclude that the adiabatic
approximation is a very good one for eq.(\ref{eq:schroed}).

The main problem one has to face with such a result, when applied to
strongly
perturbed quantum system, is the computational
difficulty that arises trying to go to higher orders. This is due to the
fact that, differently from the small-perturbation approach, the states
are time-dependent. So, the equations derived in [1] for
the strong field approximation could not be easily solvable. Beside,
our derivation does not exclude a time-independent perturbation.
But, as already
showed in [1], a constant perturbation gives rise to terms that depends
on power of time, that is, we face a small-time development. Our aim in
the
next section will be just to give an indication
that, at least in a simple case, the series
is the quantum analog of the Kirkwood-Wigner approximation of statistical
mechanics.

An interesting result we can obtain from the above discussion is that the
interaction picture, for a very large perturbation, yields the same
result as the leading order of the strong field approximation. So, let us
consider a system with a hamiltonian $H(t) = H_0 + \lambda V(t)$ with
$\lambda\rightarrow\infty$. In the interaction picture we will have
\begin{equation}
    U^+ \lambda V(t) U |\psi_I> = i\frac{d|\psi_I>}{dt} \label{eq:intpic}
\end{equation}
with
\begin{equation}
    H_0 U = i\frac{dU}{dt}
\end{equation}
and $|\psi> = U |\psi_I>$. By the conclusion drawn above, the solution of
eq.(\ref{eq:intpic}) can be written down as
\begin{equation}
    |\psi_I> \approx \sum_n c_n(t_0) e^{i\gamma_n^I (t)}
                            e^{-i\lambda\int_{t_0}^t v_n^I (t') dt'}
                            |n; t>_I
\end{equation}
being
\begin{equation}
    U^+ V(t) U |n; t>_I = v_n^I(t) |n; t>_I  \label{eq:Vstates}
\end{equation}
and
\begin{equation}
    \dot{\gamma}_n^I(t) = _I<n; t|i\frac{d}{dt}|n; t>_I. \label{eq:gammaI}
\end{equation}
It is not difficult to see that eq.(\ref{eq:Vstates}) can be rewritten as
\begin{equation}
    V(t) (U |n; t>_I) = v_n^I(t) (U |n; t>_I)
\end{equation}
that shows that $U |n; t>_I$ is an eigenstate of $V(t)$. So, using the
equation $V(t) |n;t> = v_n(t) |n; t>$, we can make the
identifications $v_n^I(t) = v_n(t)$ and
\begin{equation}
    U |n; t>_I = e^{i\alpha_n(t)}|n; t> \label{eq:Ueffect}
\end{equation}
being the phase $\alpha_n(t)$ to be determined. This can be accomplished
by
computing explicitly eq.(\ref{eq:gammaI}) using eq.(\ref{eq:Ueffect})
yielding
\begin{equation}
    \dot{\alpha}_n(t) = <n; t| iU\frac{dU^+}{dt}|n; t> = - <n; t|H_0|n; t>
\end{equation}
and $\dot{\gamma}_n^I(t) = <n; t|i\frac{d}{dt}|n; t> = \dot{\gamma}_n(t)$.
The final result is then
\begin{equation}
    |\psi> \approx \sum_n c_n(t_0) e^{i\gamma_n(t)}
                          e^{-i\int_{t_0}^t dt'
                             [<n; t'|H_0|n; t'> + \lambda v_n(t')]}
                          |n; t>.
\end{equation}
This will be a solution of eq.(\ref{eq:lo}) if the term $<n; t|H_0|n; t>$
can
be neglect respect to $v_n(t)$. This can happen by a very large
perturbation.
For an application of this result we refer back to ref.[5].

\section{The Ti\-me-In\-de\-pen\-dent Per\-tur\-ba\-tion and the
         Kirk\-wo\-od-Wig\-ner Ex\-pan\-sion}

In order to see the relation between the theory of strong perturbations
and
the Kirkwood-Wigner series of statistical mechanics, we consider a
one-dimensional particle of mass $m$
moving on a segment of lenght $L$. This particle
undergoes the effect of a potential $V(x)$, so that we consider the
Schr\"{o}dinger equation
\begin{equation}
    i\frac{\partial\psi(x,t)}{\partial t} =
    -\frac{1}{2m}\frac{\partial^2\psi(x,t)}{\partial t^2}
    + V(x) \psi(x,t)
\end{equation}
with the initial condition $\psi(x,0) = \frac{1}{\sqrt{L}}$. It is quite
easy, using the result of re.[1], to get till first order
\begin{equation}
    \psi(x,t) = \frac{1}{\sqrt{L}}
                \left\{ 1-i\frac{t^3}{6m}[V'(x)]^2
                         + \frac{t^2}{4m}V''(x)\right\}e^{-itV(x)}
\end{equation}
where we see the polynomial dependence on $t$ that makes meaningless the
series for very long time. But, if we make the substitution
$it = \beta = \frac{1}{k_B T}$, with $k_B$ the Boltzmann constant,
we recover the series in ref.[6], that is, the Kirkwood-Wigner expansion.
This is an indication that we are applying the same approximation in
quantum mechanics, giving some insight into the strongly perturbed
quantum systems. It should also be clear that a direct application of
our above results to this case can cause some difficulties.

\section{An Application: Berry's Two-level Model}

As an example, we consider the one used by Berry in [3]
\begin{equation}
    H(t) = \frac{1}{2}(X(t)\sigma_X + Y(t)\sigma_Y + Z(t)\sigma_Z)
\end{equation}
being $\sigma_X, \sigma_Y, \sigma_Z$ the Pauli matrices and now we just
retain the cyclicity of the parameters $X, Y$ and $Z$ while no adiabatic
hypothesis is made. If we have as a strong perturbation
the following part of the hamiltonian
\begin{equation}
    V(t) = \frac{1}{2}(X(t)\sigma_X + Z(t)\sigma_Z),
\end{equation}
we will fall in the same case as considered in [3],
that is, due to the above results,
the quantum system acquires a Berry's phase on a cycle given by $\pi$ and
the
wave function changes sign on a full cycle of the parameter space. This
effect could easily be put to a test.

\section{Conclusions}

We have seen that the class of strongly perturbed
system can be enlarged because,
as the perturbation becomes stronger, the quantum system approaches even
more an adiabatic one. So, the adiabatic approximation can be applied.
We have also pointed out that our approximation can be an application
to quantum mechanics of the Kirkwood-Wigner expansion of statistical
mechanics. Beside, a fairly interesting example concerns the Berry's
phase.
A geometrical phase could appear for a strongly perturbed quantum
system as in the example given in the above section.
To conclude we stress that this
enlargment of applicability of the adiabatic theory could solve
some interesting problems of field theory and quantum chaos, giving
some new and unexpected results.

\newpage

\ddag e-mail address: mc3747@mclink.it

[1] M.Frasca, Phys. Rev. A, {\bf 45}, 43 (1992);
Phys. Rev. A, {\bf 47}, 2374 (1993)

[2] A.Messiah, {\sl Quantum Mechanics}, Vol.II, North-Holland,
Amsterdam (1961); see Ch.XVII, \S 10ff.

[3] M.V.Berry, Proc. R. Soc. Lond. A {\bf 392}, 45 (1984)

[4] N.Bleistein, R.A.Handelsman, {\sl Asymptotic Expansions of Integrals},
Dover, New York, (1986)

[5] M.Frasca, Il Nuovo Cimento {\bf 109B}, 603 (1994)

[6] L.D.Landau, E.M.Lifschitz, {\sl Statistical Physics}, Pergamon Press,
Oxford (1968)

\end{document}